# Electrical Tuning of Polarizaion-state Using Graphene-Integrated Metasurfaces

Nima Dabidian[1], Shourya Dutta-Gupta[1], Iskandar Kholmanov[3,4], Mikhail Belkin[2] and Gennady Shvets[1,*]

[1]Department of Physics and Center for Nano and Molecular Science and Technology, The University of Texas at Austin, Austin, Texas 78712, United States
[2]Department of Electrical and Computer Engineering, Microelectronics Research Center, The University of Texas at Austin, 10100 Burnet Road, Austin, Texas 78758, United States.
[3]Department of Mechanical Engineering and Materials Science Program, The University of Texas at Austin, Austin, Texas 78712, United States
[4]CNR-INO, Sensor Lab, The University of Brescia, via Branze 45, 25123, Brescia, Italy
*gena@physics.utexas.edu

## Abstract

Plasmonic metasurfaces have been employed for tuning and controlling light enabling various novel applications. Their appeal is enhanced with the incorporation of an active element with the metasurfaces paving the way for dynamic control. In this letter, we realize a dynamic polarization state generator using graphene-integrated anisotropic metasurface (GIAM), where a linear incidence polarization is controllably converted into an elliptical one. The anisotropic metasurface leads to an intrinsic polarization conversion when illuminated with non-orthogonal incident polarization. Additionally, the single-layer graphene allows us to tune the phase and intensity of the reflected light on the application of a gate voltage, enabling dynamic polarization control. The stokes polarization parameters of the reflected light are measured using rotating polarizer method and it is demonstrated that a large change in the ellipticity as well as orientation angle can be induced by this device. We also provide experimental evidence that the titl angle can change independent of the ellipticity going from positive values to nearly zero to negative values while ellipticity is constant.

## Introduction

Manipulation and control of light is of extreme importance both for understanding fundamental physics as well as development of applied devices (1). Recently, metasurfaces (2D array of sub-wavelength plasmonic structures) have been developed for controlling the various properties of light like intensity, phase, polarization etc. (2,3,4,5). The choice of the unit cell as well as the lattice arrangement can be used for defining and controlling the functionality of the metasurface. In most cases, this flexibility renders the metasurfaces superior to conventional optical elements. One of the areas that metasurfaces have prooved useful in polarization tuning. Polarization conversion is instrumental to crucial optical applications such as ellipsometry, polarimetry (6), optical sensing (7) and polarization-devision multiplexing (8) which have superior performance compared to convensional methods of material characterization (9) and tellecommunication (10). This goal can be achieved by employing anisotropic metasurfaces. It has been shown that anisotropic metasurfaces can convert polarization of light from linear to circular (11,12,13,14,15,16), rotate linearly polarized light (17,18,19) and induce asymmetric tranmission (20). However, most of the studies that employed metasurfaces that are 'passive', i.e., once fabricated their properties can not be tuned or controlled. To circumvent this problem, 'active' devices are in high demand where the optical property can be tuned dynamically (5,21,22,23,24,25,26,27). The most common approaches for active tuning have been based on mechanical (28,29,30,31) or electrical stimulation (22,32). Amongst these, electrical modulation seems to allow a better speed of modulation as well as flexibility in device fabrication. At the forefront of electrical tuning, involves incorporating a material that is sensitive to electrical voltage and in this respect graphene is one of the best possibilities currently available (5,21,22,24). Such graphene integrated metasurfaces have been used for modulating the intensity as well as the phase of the reflected light (21,22,24,25,26). In fact, we recently showed that such a phase modulating device could be used as motion detector with high accuracy (26). In this paper, we show that polarization can also be controlled dynamically using a graphene integrated anisotropic metasurface (GIAM) device. A linearly polarized incident light is converted into an elliptical polarization on reflection from the anisotropic metasurface. Furthermore, applying a gate voltage to graphene allows us to control the ellipticity of the reflected light. We experimentally measure the stokes parameters of the reflected light at different gate voltage to verify this effect.

The concept used in this paper is shown in Fig. 1, where a graphene-integrated metasurface turns the polazation-state from linear (incident) to elliptical (reflected). A gate voltage is applied between a silicon back-gate and graphene to tune the ellipticity of the reflected wave.

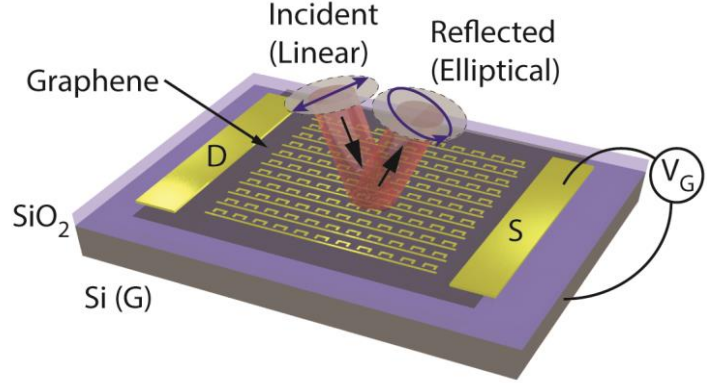

**Figure 1**: Schematic of the graphene-integrated metasurface device that converts linearly polarized incident light to elliptically polarized reflected light. The device can dynamically tune the ellipticity of the reflected wave by changing a gate-voltage applied between silicon back gate and graphene.

One possibility for inducing such a change in polarization state is via the use of anisotropic metasurfaces. The anisotropic metasurface is resonant for a given incident polarization and non-resonant for the orthogonal polarization at a given wavelength. We used a Fano-resonant plasmonic metasurface shown in Fig. 1(a) that enhances the electromagnetic field for Y-polarization. The origin of the enhancement is the Fano interference between the broadband response of the wire and the C-shaped dipole (33). At the resonance frequency, the near-fields are enhanced by $\eta = |E_t^2/E_{inc}^2|$ and an anti-symmetric current flows in the wire and the antenna as shown in Fig. 1(a). The destructive interference between these current results in a reflectivity minimum at $\lambda \approx 7.7$ μm.

The measured reflectivity from the metasurface for Y-polarized incidence is shown in Fig. 2(c) for three different gate voltages. At the charge-neutrality point (CNP) of graphene, a reflectivity minima is observed at $\lambda \approx 7.85$ μm which is a manifestation of the Fano-resonance as discussed previously. As the gate-voltage increases, the carrier concentration of graphene: $n = c\Delta V/e$ , where $c$ is the capacitance per area of the SiO$_2$ spacer and $\Delta V = V_g - V_{CNP}$ where $V_{CNP} = -200$ V for our sample. This in turn increases the Fermi energy level of graphene $E_F = \hbar v_F \sqrt{\pi n}$, where $v_F = 1*10^8$ cm/s is the Fermi velocity. As the Fermi energy increases, graphene behaves increasingly inductive, that gives rise to a blue-shifting of the resonance frequency as shown in Fig. 2(c) where the measured reflectivity $R_{yy}$ for 3 different Fermi energies are depicted. As $E_F$ increases from $E_F = 0.08$ eV at the CNP to $E_F = 0.28$ eV at $V_g$=250 V, the resonance wavelengh blueshifts from $\lambda \approx 7.85$ μm to $\lambda \approx 7.65$ μm which is around 2.6 % shift. The Fermi level $E_F$ also determines the wavelength-dependent optical conductivity $\sigma_{SLG}(\lambda, E_F)$ of graphene that we use in our simulations. The simulation results for $R_{yy}$ shown in Fig. 2(d) is in good agreement with the measured reflectivity of Fig. 2(c). The measured and simulated $R_{xx}$ are also demonstrated in Fig. 2(e),(f) which confirm that our metasurface is optically non-resonant for X-polarization at the wavelengths range of our interest. The small feature at $\lambda \approx 8$ μm in Fig. 2(e) is due to the the epsilon near zero (ENZ) effect that originates from longitudinal phonon polariton of SiO$_2$ (34,35). This effect is only excited by components of electric field that are normal to the substrate($E_z$) which are present in our experiment due to the large numerical aperture of our objective (NA=0.5). However our simulation is performed for normal incidence, thereby the ENZ feature is not present in the results as shown in Fig. 2(f). In a previous publication, we utilized an interferometric setup to experimentally measure the phase modulation for both X and Y polarized incident light (26). It was shown that the phase change as a function of gate voltage is considerable only for the light polarized along the dipole structure(Y). In this paper, we use this tunable ansiotropy to induce and measure a change in ellipticity of the light reflected from the metasurface.

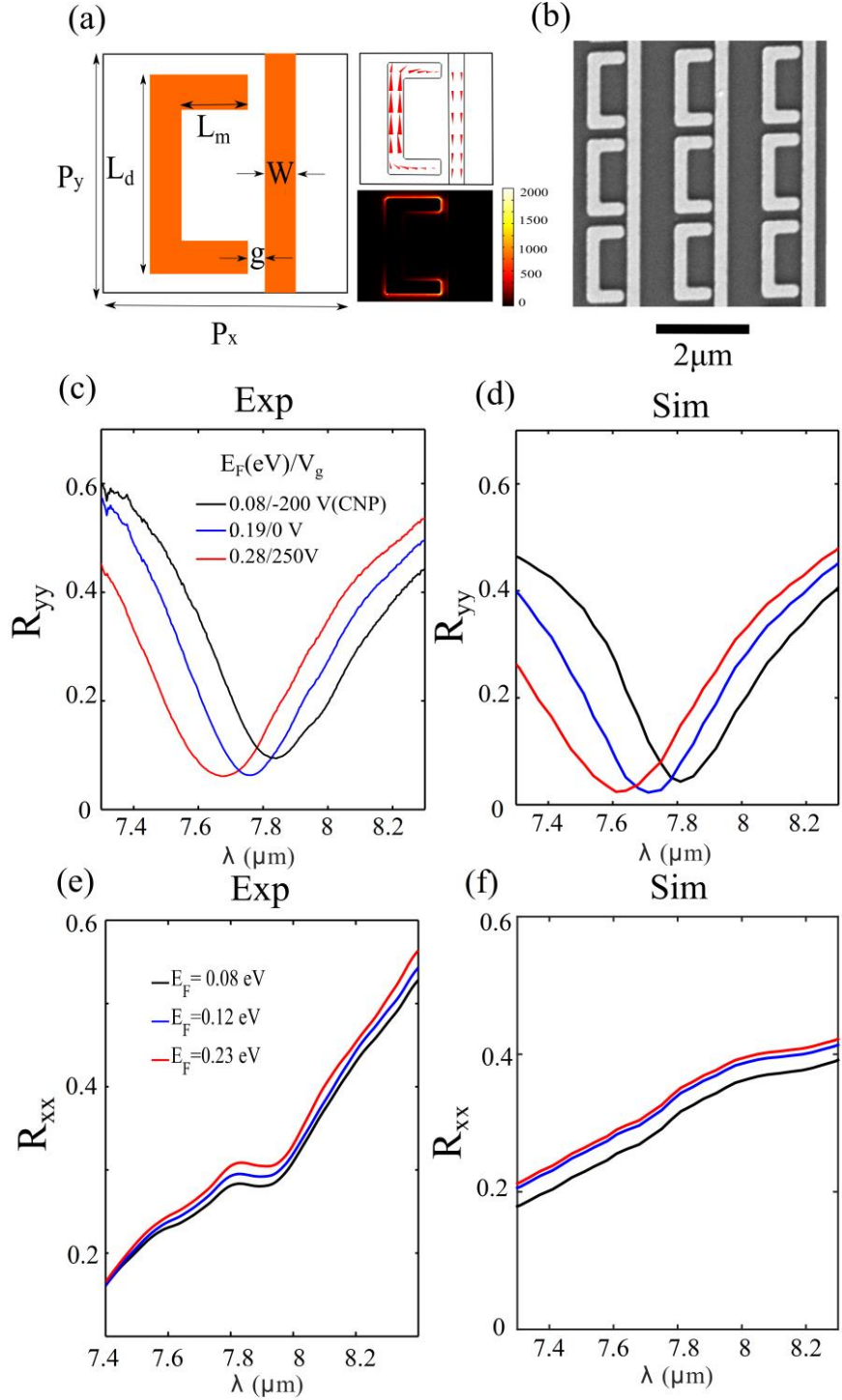

**Figure 2:** (a) The unit cell of the plasmonic metasurface with the defined structure dimensions (Left). Top right: the current density profile 5 nm above the graphene plane. Bottom right: near-field enhancement $\eta = |E_t^2/E_{inc}^2|$ at $z = 0$ surface. The field enhancement and current profiles are plotted at the resonance frequency. (b) The SEM picture of the fabricated metasurface. The black scale-bar is 2 $\mu m$ in size. (c) The measured reflectivity $R_{yy}$ spectra for different gate voltages. The legend lists the gate voltages that were applied and their corresponding Fermi energy levels. (d) The simulated reflectivity $R_{yy}$ for the Fermi levels listed in (c). (e) The measured $R_{xx}$ for the Fermi energies listed in the legend. (f) The simulated $R_{xx}$ for the Fermi levels listed in (e). The structure dimensions in (a) are the following: The periodicity of the metasurface in both directions is equal to and $P_x = P_y = 2.1$ $\mu$m. The width of all nanowires are $W = 250$ nm. Other dimensions: $g = 120$ nm, $L_d = 1.8$ $\mu$m, $L_m = 600$ nm.

In order to measure the ellipticity of the reflected light, i.e. measure the stokes parameters, we use a rotating analyzer setup shown in Fig. 3(a). The setup consists of a quantum cascade laser source of light, a polarizer, a $CaF_2$ beam-splitter, a graphene-integrated metasurface sample biased with a DC gate-voltage, an analyzer and a MCT detector. The polarization state of light after reflection from the metasurface can be described by

$$\mathbf{E} = E_x\mathbf{x} + E_y\mathbf{y} = \mathrm{Re}[E_0(\cos\theta_E\,\mathbf{x} + e^{i\phi}\sin\theta_E\,\mathbf{y})e^{-i\omega t}] \quad (1)$$

where $E_0$, $\theta_E$ and $\phi$ are the electric field amplitude, inverse tangent of amplitude ratios of $E_y$ and $E_x$ and the phase difference between $E_x$ and $E_y$, respectively. The time-averaged intensity of the analyzed signal measured at the detector then can be explained as:

$$I(E_0,\theta_E,\theta_A,\phi) = \tfrac{1}{2}E_0^2[\cos^2\theta_A\cos^2\theta_E + \sin^2\theta_A\sin^2\theta_E + \tfrac{1}{2}\sin2\theta_A\sin\theta_E\cos\phi] \quad (2)$$

where $\theta_A$ is the analyzer angle. We set the polarizer angle at $\theta_P = 75°$ which is the angle that results in equal component of X and Y polarized intensity of the reflected light at the resonance wavelength ($E_x \approx E_y$). We measured the reflection intensity as a function of wavelength by changing the analyzer angle from $\theta_A = 0°$ to $\theta_A = 90°$ in steps of $22.5°$ as shown in Fig. 3(b). A subsequent least-square fitting of equation (2) to the measured data of Fig. 3(b) at a given wavelength determines the fitting parameters $E_0, \theta_E$ and $\phi$. It is then straight forward to calculate the complex electric field of the detected signal from equation (1). The polarization state of the light can subsequently be expressed in terms of stokes parameters $S_0 = |E_x|^2 + |E_y|^2 = E_0^2$, $S_1 = |E_x|^2 - |E_y|^2 = E_0^2\cos2\theta_E$, $S_2 = 2\mathrm{Re}(E_xE_y^*) = E_0^2\sin2\theta_E\cos\phi$ and $|S_3| = -2\mathrm{Im}(E_xE_y^*) = |E_0^2\sin2\theta_E\sin\phi|$.

The normalized values of stokes parameters $\frac{S_1}{S_0}, \frac{S_2}{S_0}$ and $\frac{S_3}{S_0}$ extracted by the fitting procedure are represented in Fig. 4(a) with dashed, dotted and solid lines, respectively. The lines are color-coded according to the legend in Fig. 4(c) and similar to Fig. 2. A non-zero $S_3$ indicates an elliptical polarization of the reflected light which is expected due to the anisotropy of the metasurface. The ellipticity is changing as a function of the gate voltage e.g. at $\lambda = 7.76$ μm tagged by the magenta dashed line, $S_3 = S_0$ for zero gate-voltage $V_g = 0$ V which means the light is nearly circularly polarized. For the other two voltages, $V_g = -200$ V and $V_g = 250$ V, light will have elliptical polarization with different ellipticities.

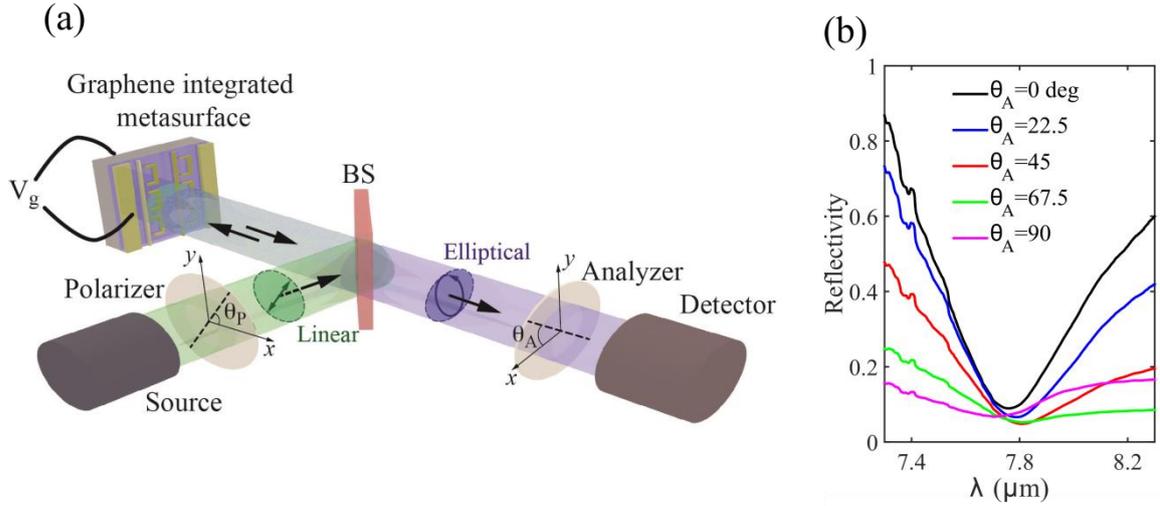

**Figure 3**: (a) The schematic for the stokes polarimetry setup consisting of an optical source, a beam splitter (BS), an infrared MCT detector, two linear polarizers: one after the source and one before the detector that is used as an analyzer of the polarization state. A graphene-integrated metasurface device that can alter the polarization state of the incident light by applying a gate voltage $V_g$ is incorporated into the setup. (b) The reflectivity as a function of wavelength for different analyzer angles when the structure is illuminated with light polarized at $\theta_P=75^0$.

Figure 4(c) shows the polarization ellipses at $\lambda = 7.76$ μm for the three different gate voltages. Polarization ellipse, as schematically shown in Fig. 4(b), is characterized by its ellipticity $\frac{a}{b}$ and tilt angle β, where a and b are the semi-major axis and semi-minor respectively and $\tan2\beta = S_2/S_1$. As the gate voltage increases from $V_g = -200$ V to $V_g = 0$ V, the polarization state changes from elliptical with ellipticity of $\frac{a}{b} \approx 2.1$ to almost circular ($\frac{a}{b} \approx 1$) as shown by the blue circle. Further increasing the gate voltage from $V_g = 0$ V to $V_g = 250$ V the ellipticity increases from 1 to 4.5. The polarization ellipse at the gate voltage of $V_g = 250$ V has a larger tilte angle of $\beta \approx -37°$ compared to $V_g = -200$ V with $\beta \approx -16°$.

At another interesting wavelength of λ = 7.72 μm tagged by the green dashed line in Fig 4(a), the third component of normalized stokes parameters $\frac{S_3}{S_0}$ are identical for the three gate voltages rendering equal ellipticity. As the voltage increases from $V_g = -200$ V to $V_g = 250$ V the polarization ellipse rotates clockwise and the tilt angle changes sign as it goes from β ≈ 31° to β ≈ −41° as depicted in Fig 4(d). At $V_g = 0$ V the tilt angle is small: β ≈ −6°. Therefore the tilt angle is changing while ellipticity is kept constant. Such independent modulation of tilt angle from ellipticity, can be useful in polarization-devision multiplexing (36).

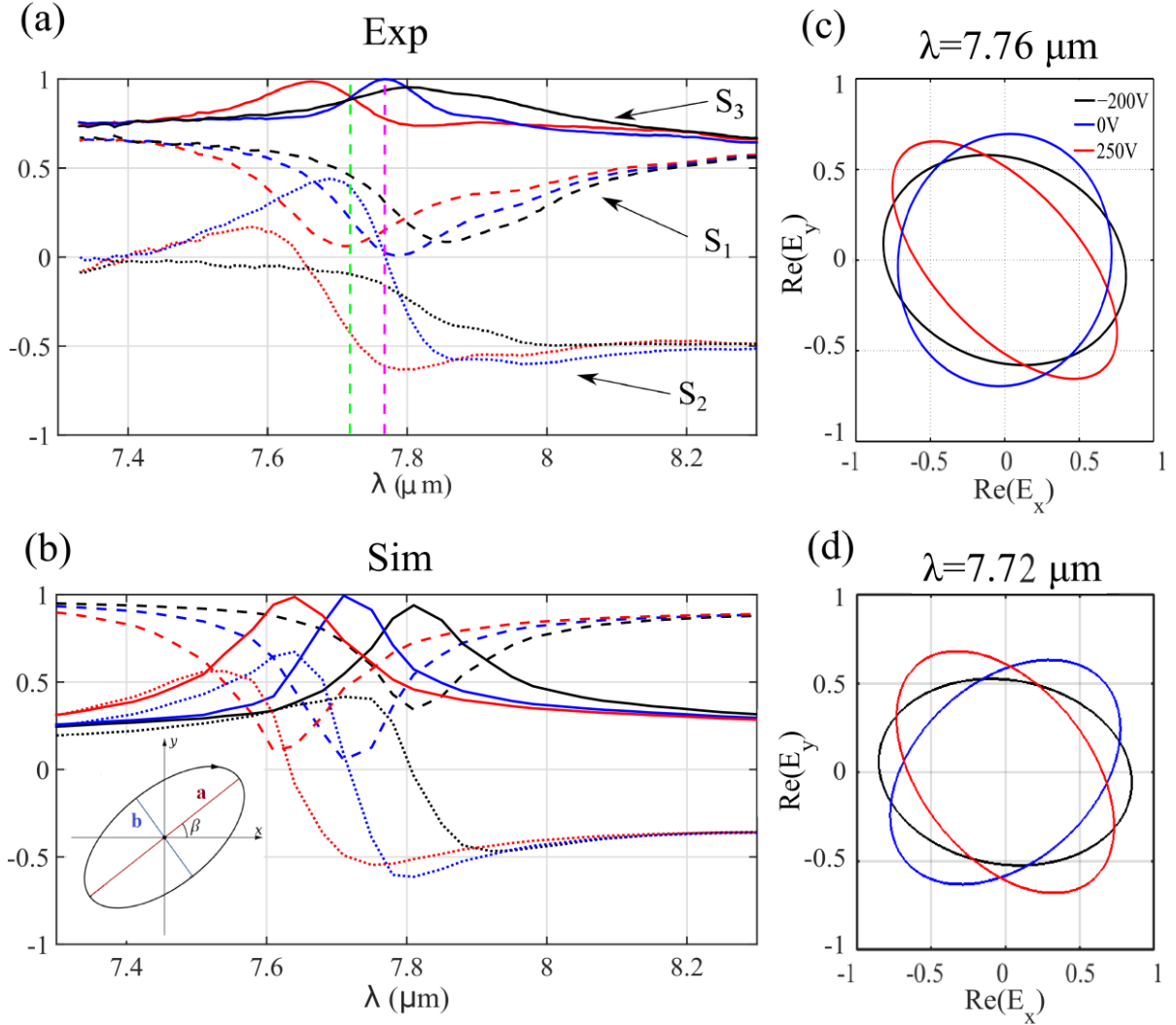

**Figure 4:** (a) Measured and (b) simulated determined stokes parameters of the reflected light as a function of wavelength. (c) and (d) The polarization ellipses at λ = 7.76 μm and λ = 7.72 μm for three different gate voltages.

To confirm the experimentally measured stokes parameters, we performed numerical simulations to evaluate the ellipiticity of the light reflected from the metasurface when illuminated at normal incidence and with light polarized along $θ_p = 75°$. The calculated stokes parameters as a function of wavelength are shown in Fig. 4(b) which is in good agreement with the measured results of Fig 4(a).

# Conclusions

In this paper, we provided experimental evidence for active modulation of polarization-state of light reflected from a graphene integrated anisotropic metasurface. The rotating analyzer approach was used to experimentally determine the stokes parameters of the reflected light. Polarization ellipses drawn at representative wavelengths for the different gate voltages clearly show that we are able to controllably change the polarization state of the light from linear to elliptical or circular. In general, it is shown that both the tilt angle and ellipticity can be changed on applying a gate voltage. Furthermore, it is shown that the polarization ellipse can rotated from 1st to the 4th quadrant while keeping its ellipticity constant. This modulation of tilt angle, independent of the ellipticity, can be used for polarization encoding of light and could be instrumental for improving the telecommunication bandwidth.

# Methods

*Optical simulations*: We used a commercial finite elements solver COMSOL Multiphysics version 4.3b to simulate the reflectivity of the sample. To model SLG, a surface current (37) $J_{SLG} = \sigma_{SLG} E_t$ was defined where $E_t$ is the tangential electric field on the SLG plane and $\sigma_{SLG}$ is the optical conductivity of graphene that was calculated from random-phase approximation (38):

$$\sigma_{inter}(\omega) = \frac{e^2}{4\hbar}\left[\frac{1}{2} + \frac{1}{\pi}\tan^{-1}\left(\frac{\hbar\omega - 2E_F}{2k_B T}\right) - \frac{i}{2\pi}\ln\frac{(\hbar\omega + 2E_F)^2}{(\hbar\omega - 2E_F)^2 + (2k_B T)^2}\right],$$
$$\sigma_{intra}(\omega) = \frac{e^2}{4\hbar}\frac{8ik_B T}{\pi\hbar(\omega + i\tau^{-1})}\ln\left[2\cosh\left(\frac{E_F}{2k_B T}\right)\right]$$

Where $k_B$ is the Boltzmann constant. In our calculation, we assumed room temperature (T = 300 K) and the carrier scattering time of $\tau = 13$ fs was used.

*Sample fabrication*: SLG was grown on polycrystalline Cu foil using a CVD technique (39) and then transferred (40) onto a commercially purchased Si/SiO$_2$ substrate (University Wafer) with 1 µm oxide layer grown on lightly doped silicon. Elecon-beam patterning followed by an oxygen plasma cleaning step isolated the high-quality graphene regions. The anisotropic metasurface with unit cell of Fig. 2(a) was fabricated by e-beam lithography on top of SLG in area of 100 µm × 100 µm. The thickness of the metasurface was 30 nm (5 nm of Cr and 25 nm of Au). Source and drain contacts of 100 nm thick (15 nm Cr + 85 nm Au) were deposited on top of graphene on the two sides of the metasurface samples. Finally, a metallic contact (15 nm Ni+85 nm Au) was deposited on the back of the silicon sample for gating. Wirebonding of the contacts to a chip-carrier concluded the fabrication.

*Reflection measurement*: The setup shown in Fig. 3(a) was used to measure the optical reflectivity of the sample. The source was a quantum cascade laser (Daylight solution, MIRcat-1400) operated in pulsed mode with the pulse repetitions rate of 250 KHz and the pulse duration of 100 ns. A ZnSe objective lens with high numerical aperture (NA=0.5) was utilized to focus the laser light onto the metasurface. A liquid-nitrogen-cooled mercury-cadmium-telluride (MCT) detector measured the intensity of the signal. The measured signal was utilized for the measurements of signal intensity. A lock-in amplifier (Stanford research systems SR844) was used to amplify the measured signal with a 3 ms integration time.

# References

1. Born M, Wolf E. Principles of Optics: Cambridge University Press; 1999.
2. Kildishev AV, Boltasseva A, Shalaev VM. Planar photonics with metasurfaces. *Science*. 2013; 339(6125): p. 6125.
3. Meinzer N, Barnes WL, Hooper IR. Plasmonic meta-atoms and metasurfaces. *Nature Photonics*. 2014; 8: p. 889-898.
4. Yu N, Capasso F. Flat optics with designer metasurfaces. *Nature Materials*. 2014; 13: p. 139-150.
5. Minovich AE, Miroshnichenko AE, Bykov AY, Murzina TV, Neshev DN, Kivshar YS. Functional and nonlinear optical metasurfaces. *Laser & Photonics Reviews*. 2015; 9(2): p. 195-213.
6. Losurdo M, Bergmair M, Bruno G, Cattelan D, Cobet C, De Martino A, et al. Spectroscopic ellipsometry and polarimetry for materials and systems at the nanometer scale: State-of-the-Art, Potential, and Perspectives. *J. Nanoparticle Res.* 2009; 11(7): p. 1521-1554.
7. Adato R, Yanik A. A. , Ansden JJ, Kaplan DL, Omenetto FG, Hong MK, et al. Ultra-sensitive vibrational spectroscopy of protein monolayers with plasmonic nanoantenna arrays. *Proceedings of the National Academy of Science.* 2009; 106(46): p. 19227-19232.
8. Morant M, Rez J, Llorente R. Polarization division multiplexing of OFDM radio-over-fiber signals in passive optical networks. *Adv. Opt. Technol.* 2014; 2014(e269524).
9. Zamani M, Shafiei F, Fazeli SM, Downer MC, Jafari GR. Analytic heigh correlation function of rough

surfaces derived from light scattering. *AsXiv Prepr. 2015*;(ArXiv150700434).
10. Miller S. Optical fiber telecommunications: Elsevier; 2012.
11. Hao J, Yuan Y, Ran L, Jiang T, Kong JA, Chan CT, et al. Manipulating electromagnetic wave polarizations by anisotropic metamaterials. *Phys. Rev. Lett.* 2007; 99(6): p. 063908.
12. Pors A, Nielsen MG, Valle GD, Willatzen M, Albrektsen O, Bozhevolnyi SI. Plasmonic metamaterial wave retarders in reflection by orthogonally oriented detuned electrical dipoles. *Opt. Lett.* 2011; 36(9): p. 1626.
13. Wang F, Chakrabarty A, Minkowski F, Sun K, Wei EH. Polarization conversion with elliptical patch nanoantennas. *Appl. Phys. Lett.* 2012; 101(2): p. 023101.
14. Khanikaev AB, Mousavi SH, Wu C, Dabidian N, Alici KB, Shvets G. Electromagnetically induced polarization conversion. *Opt. Commun.* 2012; 285(16): p. 3423-3427.
15. Abasahl B, Dutta-Gupta S, Santschi C, Martin OJF. Coupling Strength Can Control the Polarization Twist of a Plasmonic Antenna. *Nano Lett.* 2013; 13(9): p. 4575–4579.
16. Li Y, Zhang J, Qu S, Wang J, Zheng L, Pang Y, et al. Achieving wide-band linear-to-circular polarization conversion using ultra-thin bi-layered metasurfaces. *J. Appl. Phys.* 2015; 117(4): p. 044501.
17. Hao J, Ren Q, An Z, Huang X, Chen Z, Qiu M, et al. Optical metamaterial for polarization control. *Phys. Rev. A.* 2009; 80(2): p. 023807.
18. Dai Y, Ren W, Cai H, Ding H, Pan N, Wnag X. Realizing full visible spectrum metamaterial half-wave plates with patterned metal nanoarray/insulator/metal film structure. *Opt. Express.* 2014; 22(7): p. 7465-7472.
19. Chen H, Wang J, Ma H, Qu S, Xu Z, Zhang A, et al. Ultra-wideband polarization conversion metasurfaces based on multiple plasmon resonances. *J. Appl. Phys.* 2014; 115(15): p. 154504.
20. Fedotov VA, Mladyonov PL, Prosvirnin SL, Rogacheva AV, Chen Y, Zheludev NI. Asymmetric propagation of electromagnetic waves through a planar chiral structure. *Phys. Rev. Lett.* 2006; 97(16): p. 167401.
21. Yao Y, Kats MA, Genevet P, Yu N, Song Y, Kong J, et al. Broad Electrical Tuning of Graphene-Loaded Plasmonic Antennas. *Nano Letters.* 2013; 13(3): p. 1257-1264.
22. Yao Y, Kats MA, Shankar R, Song Y, Kong J, Loncar M, et al. Wide Wavelength Tuning of Optical Antennas on Graphene with Nanosecond Response Time. *Nano Letters.* 2014; 14(1): p. 214-219.
23. Si G, Zhao Y, Leong E, Liu Y. Liquid-crystal-enabled active plasmonics: A review. *Materials.* 2014; 7(2): p. 1296.
24. Emani NK, Chung TF, Kildishev AV, Shalaev VM, Chen YP, Boltasseva A. Electrical modulation of fano resonance in plasmonic nanostructures using graphene. *Nano Letters.* 2014; 14(1): p. 78-82.
25. Dabidian N, Kholmanov I, Khanikaev AB, Tartar K, Tredafilov S, Mousavi SH, et al. Electrical Switching of Infrared Light Using Graphene Integration with Plasmonic Fano Resonant Metasurfaces. *ACS Photonics.* 2015; 2(2): p. 216-227.
26. Dabidian N, Dutta-Gupta S, Kholmanov I, Lai K, Lu F, Lee J, et al. Experimental Demonstration of Phase Modulation and Motion Sensing Using Graphene-Integrated Metasurfaces. *Nano Letters.* 2016; 16(6): p. 3607-3615.
27. Zheludev NI, Plum E. Reconfigurable nanomechanical photonic metamaterials. *Nat. Nano.* 2016; 11(1): p. 16-22.
28. Huang F, Baumberg JJ. Actively tuned plasmons on elastomerically driven Au nanoparticle dimers. *Nano Letters.* 2010; 10(5): p. 1787-1792.
29. Shrekenhamer D, Chen WC, Padilla WJ. Liquid crystal tunable metamaterial perfect absorber. *Phys. Rev. Lett.* 2013; 110: p. 177403.
30. Kim JT. CMOS-compatible hybrid plasmonic modulator based on vanadium dioxide insulator-metal phase transition. *Opt. Lett.* 2012; 39(13): p. 3997.
31. Peng F, Chen H, Tripathi S, Twieg RJ, Wu ST. Fast-response IR spatial light modulators with a polymer network liquid crystal. *In SPIE Proceedings;* 2015. p. 93840N-8.
32. Jun YC, Brener I. Electrically tunable infrared metamaterials based on depletion-type semiconductor devices. *J. Opt.* 2012; 14(11): p. 114013.
33. Zhang S, Genov DA, Wang Y, Liu M, Zhang X. Plasmon-induced transparency in metamaterials. *Phys. Rev. Lett.* 2008; 101(4): p. 047401.

34. Chen DZA, Chen G. Measurement of silicon dioxide surface phonon-polariton propagation length by attenuated total reflection. *Appl. Phys. Lett.* 2007; 97(12): p. 121906.

35. Vasant S, Archmanbault A, Marquier F, Pardo F, Gennser U, Cavanna A, et al. Epsilon-near-zero mode for active optoelectronic devices. *Phys. Rev. Lett.* 2012; 109(23): p. 237401.

36. Gopinathan U, Naughton TJ, Sheridan JT. Polarization encoding and multiplexing of two dimensional signals: Application to image encryption. *Appl. Opt.* 2006; 45(22): p. 5693-5700.

37. Mousavi SH, Kholmanov I, Alici KB, Purtseladze D, Arju N, Tartar K, et al. Inductive tuning of fano-resonant metasurfaces using plasmonic response of graphene in the mid-infrared. *Nano Letters.* 2013; 13(3): p. 1111-1117.

38. Falkovsky LA, Pershoguba SS. Optical far-infrared properties of a graphene monolayer and multilayer. *Phys. Rev. B*. 2007; 76(15): p. 153410.

39. Li X, Zhu Y, Cai W, Borysiak M, Han B, Chen D, et al. Transfer of large-area graphene films for high-performance transparent conductive electrodes. *Nano Letters.* 2009; 9(12): p. 4359-4363.

40. Suk JW, Kitt A, Magnuson CW, Hao Y, Ahmed S, An J, et al. Transfer of CVD-grown monolayer graphene onto arbitrary substrates. *ACS Nano.* 2011; 5(9): p. 6916-6924.